\documentclass[runningheads]{llncs}

\usepackage{amsmath,amssymb,amsfonts}
\usepackage{algorithm}
\usepackage{algpseudocode}
\usepackage{graphicx}
\usepackage{textcomp}
\usepackage{xcolor}
\usepackage{tabularx}
\usepackage{booktabs}
\usepackage{url}
\usepackage{subcaption}
\usepackage[skip=1pt]{caption}
\usepackage{makecell}  % in preamble
\usepackage{float} % for [H]
\usepackage{url} 
\usepackage{caption}
\usepackage{subcaption}
\usepackage{multirow}
\usepackage{hyperref} 
\usepackage{caption}
\usepackage{chngcntr}
\usepackage{etoolbox}
\usepackage{multicol}

\AtBeginEnvironment{equation}{\setlength{\abovedisplayskip}{5pt}\setlength{\belowdisplayskip}{5pt}}
\AtBeginEnvironment{equation*}{\setlength{\abovedisplayskip}{3pt}\setlength{\belowdisplayskip}{3pt}}

\usepackage[T1]{fontenc}
\usepackage{graphicx}
\begin{document}
\titlerunning{Schema-Agnostic Process Trace Construction}
\title{Schema-Agnostic Process Trace Construction: From Raw Tables to Execution Behavior}
\author{Zhi Quan Joel Lim\inst{1}\orcidID{0009-0000-3756-7439} \and
Kar Way Tan\inst{1}\orcidID{0000-0002-2707-6588} \and
Hoong Chuin Lau\inst{1}\orcidID{0000-0002-5326-411X} }
\authorrunning{Lim et al.}
\institute{
\textsuperscript{1}School of Computing and Information Systems, Singapore Management University, Singapore\\
\email{\{joellim.2022, kwtan, hclau\}@smu.edu.sg}
}
\maketitle              % typeset the header of the contribution
\begin{abstract}

Traditional information systems (IS) engineering assumes stable schemas, explicit keys, and curated event logs. In modern OLTP environments, schemas drift, keys are sparse, and execution traces are dispersed across loosely connected tables, making manual process trace construction costly and error prone. We propose a schema-agnostic pipe-line that automatically reconstructs process execution traces directly from raw relational data. The pipeline (i) identifies columns that function like keys or timestamps, (ii) discovers table-to-table connections using statistical signals rather than predefined schemas, (iii) assembles and orders events for each case while accommodating multiple date fields, and (iv) learns likely ordering and flow relations across systems using a Temporal Convolutional Network which models long-range dependencies and patterns. Evaluations on TPC-H/E benchmarks, synthetic corpora, and a real industry dataset show that our pipeline reconstructs high-fidelity event traces and accurate trace orderings, correctly predicting the next event with 85\% accuracy and recovering about 82\% of ground-truth precedence relations. By eliminating dependence on predefined schemas, ER diagrams and domain templates, our work offers a generalizable and scalable pathway for automated reconstruction of execution behaviour in dynamic and continuously evolving IS environments.
1

\keywords{Process Mining  \and Event Log Generation \and Temporal Convolutional Network \and Financial Information Systems \and Data Profiling}
\end{abstract}

\section{Introduction}

Modern information systems (IS) increasingly rely on complex, rapidly evolving OLTP infrastructures where business activities are executed across loosely coupled relational tables. In these environments, however, schemas drift over time, surrogate keys are introduced inconsistently and are refactored into lookup tables or omitted, and execution traces are scattered across loosely coupled systems. As a result, reconstructing system behaviour requires manual joins over key‑sparse tables and ad hoc timestamp rules, which is costly to build and fragile to maintain. This bottleneck has motivated the search for scalable, domain-agnostic solutions that can automatically infer execution behaviour or operational traces from raw OLTP data by identifying inter-table relationships and data flows without relying on predefined schemas or domain expertise \cite{10.4018/978-1-60566-669-3.ch023,10.1007/978-3-030-58666-9_29,10.2308/horizons-2022-153}.

Recovering these execution traces is crucial for a broad range of IS tasks including auditing, compliance monitoring, system modernization, and the generation of event-centric data for downstream applications such as process mining. However, these possibilities are hindered by the substantial effort required to assemble high-quality execution sequences at scale. Evidence of process execution is dispersed across systems. Suitable case identifiers are difficult to determine and multiple timestamp fields often coexist within and across tables, often with ambiguous semantics. As organizations introduce new applications and integrations, the effort required to maintain execution-trace reconstruction pipelines grows disproportionately, underscoring the need for scalable and schema-agnostic approaches. 
\vspace{-0.6em}
\paragraph{Problem Statement:} Given a collection of heterogeneous, key-sparse relational tables from evolving OLTP systems, construct high-fidelity execution sequences, including cases, events, and their temporal ordering, but without relying on predefined schemas, reliable foreign keys, or domain-specific templates.

Existing approaches generally rely on explicitly declared and reliable PK/FK structures, well-defined timestamp semantics, or domain knowledge. Such assumptions rarely hold, limiting applicability in continuously changing environments. 

\begin{figure}[!t]
    \includegraphics[width=0.92\textwidth]{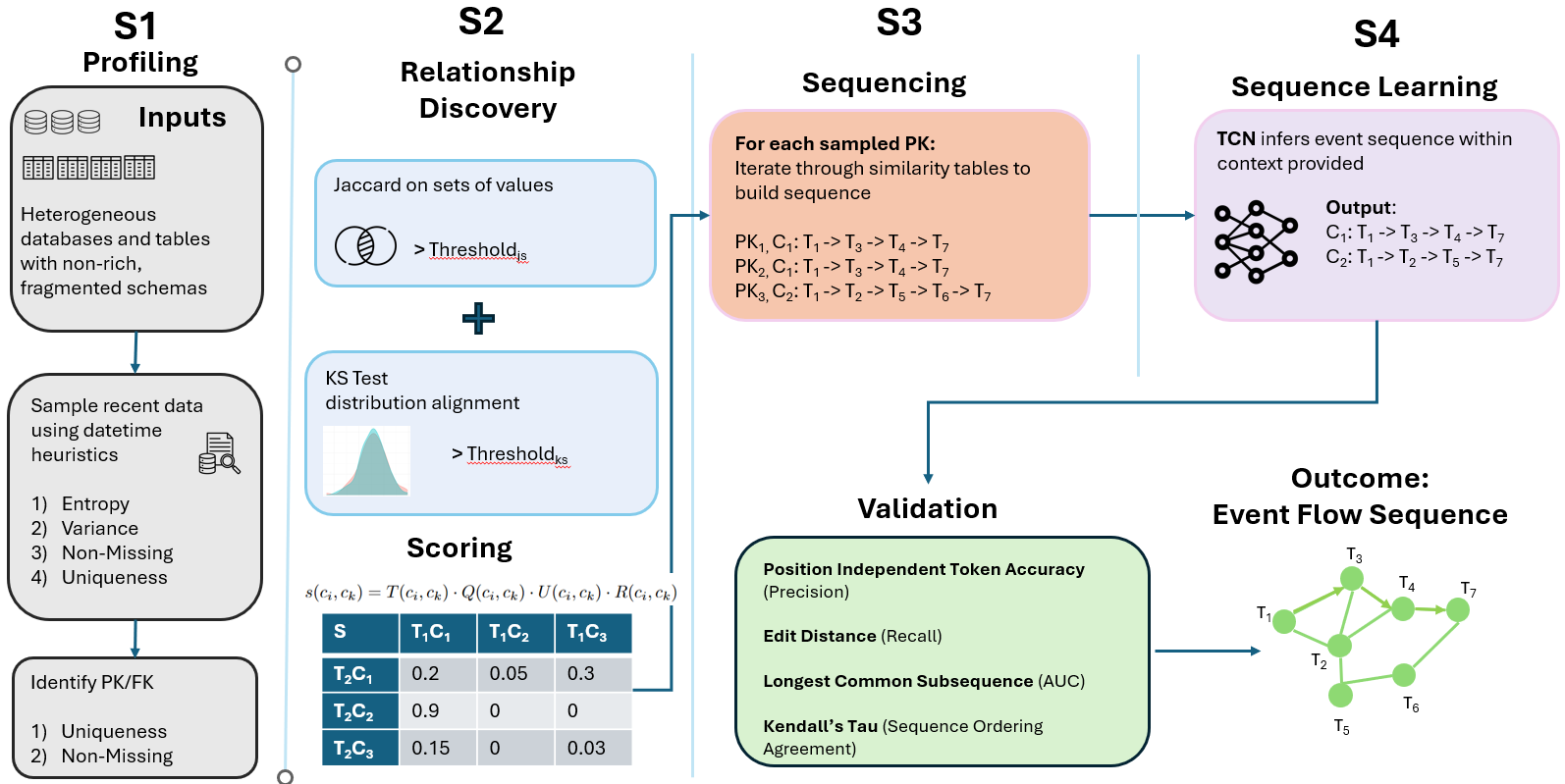}
    \caption{Overview of our end-to-end methodology for reconstructing event flows from heterogeneous, schema-poor databases.}
    \label{fig:Method}
\end{figure}

In this paper, we propose a schema-agnostic pipeline that reconstructs process events directly from raw relational data. Our approach is designed for evolving, multi-system landscapes, as it relies on the data itself rather than predefined schemas, tolerates sparsity and noise, and learns likely precedence and flow relations information when timestamps are ambiguous, partially missing or insufficient to determine the process execution order. Concretely, the pipeline (see Figure \ref{fig:Method}): 
(i) identifies columns that function as case identifiers or timestamps,
(ii) discovers relationships between tables using statistical indicators rather than explicit foreign keys,
(iii) assembles and orders events for each case while accommodating multiple timestamp fields, and
(iv) learns likely precedence and flow relations across systems using a Temporal Convolutional Network (TCN). The resulting traces can be consumed directly by standard process-mining tools and can be lifted to object-centric representations when object relations are available, ensuring compatibility without relying on domain-specific templates.

Through this work, we address the following research questions (RQ):
\begin{enumerate}
    \item \textbf{RQ1} (Event Recovery): How accurately can our pipeline recover the correct set of execution events from heterogeneous, key-sparse relational databases?
    \item \textbf{RQ2} (Precedence / Flow Learning): Can a TCN learn cross-table precedence relations more accurately than rule based approaches?
    \item \textbf{RQ3} (Change Resilience and Scalability): How resilient is our approach to noise, missing data, timestamp-semantics variation, and data-volume growth that characterize continuously changing information environments?
\end{enumerate}
In our approach, we learn cross table precedence relations without relying on fixed schemas while maintaining performance under drift. Across synthetic, benchmark, and industry datasets, the approach consistently recovers the large majority of events that belong to each process (typically over 90\%), reconstructs event sequences with high fidelity, and captures cross-table flow relations with accuracy exceeding 80\%. These results demonstrate the robustness and generalizability of the proposed approach.

Our contribution is both methodological and operational. Methodologically, we avoid the traditional requirement for complete schemas, explicit primary and foreign keys, or ontological mappings before log construction. Instead, we infer only the high confidence links needed to assemble reliable traces, and we use learned temporal patterns to order events when timestamps are insufficient. Operationally, we provide an end-to-end pathway from raw and evolving relational data to provenance-rich event logs that remain robust to missing timestamps, homonyms, and partial keys, and that can be  consumed directly by existing process mining tools. By replacing schema-first assumptions with data-driven similarity and learned precedence, the approach enables automated and reproducible log construction in environments where structures and data meaning change continuously.

\section{Related Work}
A recent systematic review of 260 process-mining studies highlights that most effort remains focused on pre-analysis and data acquisition, while scalable end-to-end log construction from heterogeneous databases is still underexplored \cite{10.1145/3712587}. This motivates automated approaches for key-sparse, legacy warehouses of the type addressed in this work.

Artifact-centric methods reconstruct primary/foreign-key structure before mapping artifacts to logs \cite{doi:10.1142/S021884301550001X}. Automated variants \cite{10.1007/978-3-642-36285-9_36} infer column types, keys, foreign keys, and timestamp attributes, then cluster tables into artifact schemas. These techniques assume that PK/FK-like patterns and a suitable timestamp can be reliably recovered. When schemas are clean and stable, \cite{10.1007/978-3-642-36285-9_36} is appropriate, otherwise, schema-agnostic pipelines that tolerate key sparsity and multiple candidate timestamp remain necessary.

Extensions mine cross-artifact interactions via graph searches \cite{7229358}, while semi-automated or constraint-based extractors additionally rely on domain inputs, predefined keys, or ontology-guided mappings or process documents \cite{ANDREWS2020113265,calvanese2016ontology,10.1007/978-3-031-95397-2_7}. Hierarchy-discovery methods require existing logs and deterministic key mining \cite{10.1007/978-3-319-10172-9_7}, and meta-model approaches depend on explicit inter-table dependencies \cite{10.1007/s10270-018-0664-7}. Legacy warehouses, however, often lack declared keys or exhibit only approximate inclusions \cite{app122110832}, making these approaches difficult to operationalize. Specialized pipelines exist for platforms such as SAP-centric OCEL extraction \cite{10.1007/s10844-023-00799-9}, but they do not generalize to heterogeneous environments. Our focus instead is on constructing log-ready linkages directly from raw tables without full schema reconstruction or domain knowledge.

Schema profiling research provides strong techniques for discovering UCC, INDs and PK/FK candidates \cite{10.1007/s10844-019-00562-z}, as well as surveys of profiling operators relevant for large-scale relationship inference \cite{10.1007/s00778-015-0389-y} and pipelines for schema normalization \cite{10.5441/002/edbt.2017.31}. We draw on these insights but restrict our objective to \emph{log-ready} linkage discovery: we compute JS/KS-based similarity on sampled data, retain only high-confidence edges, and directly assemble event sequences rather than performing complete schema recovery. Unlike profiling-only or template-based extractors, our method is designed to operate under key sparsity, multiple timestamp fields, and noise.

Some techniques assume complete event logs and model uncertainty on top \cite{10680678}, while others infer sequences from raw databases using predefined causal process templates \cite{Waibel2022CausalPM}. Synthetic-log frameworks and redo-log mining \cite{10.1007/978-3-031-16103-2_14,10.52825/bis.v1i.66} are useful when process models or vendor traces exist, but rely on templates or proprietary infrastructure often unavailable in finance. Complementary work on process-data quality highlights issues such as timestamp gaps, inconsistent labels, and incomplete keys \cite{10.1145/3613247}, which our pipeline mitigates through column profiling, similarity-based filtering, and learned ordering.

Predictive monitoring over \emph{existing} logs typically employs recurrent networks \cite{10.1007/978-3-319-59536-8_30}. Temporal Convolutional Networks (TCNs) provide long receptive fields and causal convolutions suited for irregular enterprise sequences \cite{Bai2018AnEE}. We instead use TCNs to \emph{learn ordering itself} across events reconstructed from raw relational data—distinct from work that assumes a ready-made log.

In summary, while methodologies such as OCPM$^2$ offer a valuable methodological blueprint for OCED extraction\cite{10.1007/978-3-031-95397-2_8}, they and prior work still presuppose substantial manual configuration by either reconstructing schemas before logging \cite{10.1007/978-3-642-36285-9_36,doi:10.1142/S021884301550001X}, assume reliable keys or existing logs \cite{ANDREWS2020113265,10.1007/978-3-319-10172-9_7,10.1007/s10270-018-0664-7,10680678}, rely on templates or vendor traces \cite{Waibel2022CausalPM,10.1007/978-3-031-16103-2_14,10.52825/bis.v1i.66}, or focus on profiling without end-to-end log generation \cite{10.1007/s10844-019-00562-z,10.1007/s00778-015-0389-y,10.5441/002/edbt.2017.31}. In contrast, we contribute a schema-agnostic pipeline that infers high-confidence inter-table links, selects timestamp fields automatically, and learns cross-system ordering with TCNs, addressing the gap identified in \cite{10.1145/3712587} and the quality challenges noted in \cite{10.1145/3613247}, while complementing platform-specific extractors \cite{10.1007/s10844-023-00799-9}.

\section{Methodology}

We propose a schema-agnostic pipeline that infers inter table links from statistical signals, assembles temporal fields into coherent event sequences, and uses a causal TCN to learn ordering, as shown in Figure \ref{fig:Method}. The pipeline consists of four stages that correspond to the subsections below:
\begin{itemize}
    \item \textbf{S1 Profiling Identifier and Timestamp Columns:} Detects identifier-like and temporal attributes through column profiling.
    \item \textbf{S2 Relationship Discovery:} Discovers inter-table links using type-aware statistical similarity measures.
    \item \textbf{S3 Sequencing:} Assembles timestamped rows into ordered event sequences by propagating sampled identifiers.
    \item \textbf{S4 Sequence Learning:} Learns context-dependent precedence relations using a TCN.
\end{itemize}

Stages S1–S3 produce the event traces while S4 summarizes these traces into precedence and flow graphs without modifying their event order.
Although several components of our pipeline (e.g., identifier scoring, timestamp ranking, similarity thresholds) use lightweight heuristics, these choices are principled rather than ad hoc. They operate on minimal statistical signals such as distinctness, completeness, stability, and value-overlap, which are the few reliable indicators in schema-poor, key-sparse environments. The TCN learns precedence and flow patterns directly from reconstructed traces, capturing cross-table ordering even when timestamps conflict or exhibit drift. This overall design enables the construction of provenance-rich traces that can be consumed directly by process mining tools and lifted to object-centric representations.

\subsection{Profiling Identifier and Timestamp Columns}
\begin{figure}[H]
  \centering
    \includegraphics[width=0.60\textwidth]{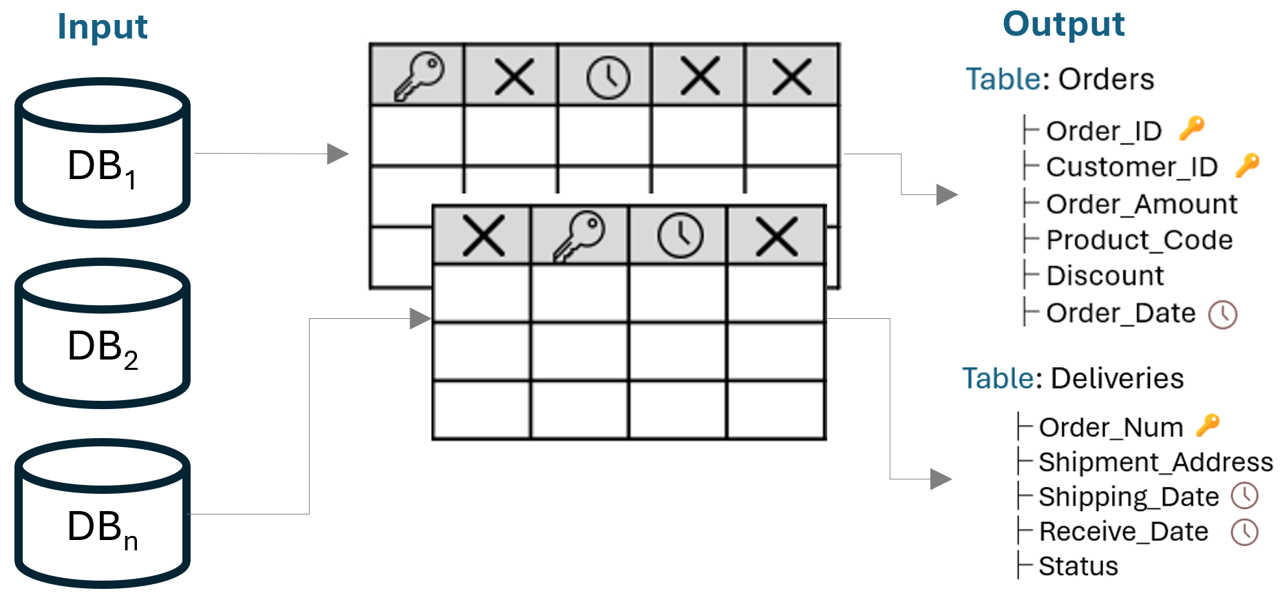}
    \caption{Profiling Identifier and Timestamp Columns: Computes identifier and datetime likelihood per column.}
    \label{fig:ProfilePrimitives Summary}
\end{figure}

We first profile each column $c$ in table $T_i$ to estimate its likelihood of acting as an identifier and its suitability as a timestamp (Figure~\ref{fig:ProfilePrimitives Summary}). Identifier likelihood combines three simple statistics:
\[
s_{\mathrm{pk}}(c) = u(c)\,\gamma(c)\,\lambda(c),
\]
where $u(c)$ is distinctness  
% \(
% u(c)=\frac{|\mathrm{distinct\_nonnull}(c)|}{|\mathrm{nonnull}(c)|},
% \)
$\gamma(c)$ is completeness and $\lambda(c)$ measures string-length stability. Columns with high $s_{\mathrm{pk}}$ are retained as candidate case identifiers. We require all three properties to be high simultaneously as multiplicative combination penalizes columns that are strong in only one dimension.

% \(
% \gamma(c)=\frac{|\mathrm{nonnull}(c)|}{|c|},
% \)
% \(
% \lambda(c)=\frac{1}{1 + \mathrm{CV}(\mathrm{strlen}(c))}.
% \)

\[
s_{\mathrm{dt}}(c)
= w_{\gamma}\pi_{\gamma}(c)
+ w_{\delta}\pi_{\delta}(c)
+ w_{\mathrm{sd}}\pi_{\mathrm{sd}}(c)
+ w_{\mathrm{IQR}}\pi_{\mathrm{IQR}}(c)
+ w_{H}\pi_{H}(c),
\]
where $\pi_{*}$ denote the per-table percentile ranks of the corresponding dispersion and variation metrics (completeness $\gamma$, distinctness $\delta$, standard deviation, inter-quartile range, and entropy)\footnote{In all experiments, we assign equal weights
\((w_{\gamma},\, w_{\delta},\, w_{\mathrm{sd}},\, w_{\mathrm{IQR}},\, w_{H})\) 
to the five percentile-based dispersion metrics. This avoids dataset-specific tuning and keeps the profiling stage parameter-light. In practice, we observed that moderate variations in these weights did not materially change the ranking of timestamp candidates.}. Per table, we retain the top-$k$ identifier and timestamp columns. These candidate identifiers and timestamp columns form the basis for subsequent link discovery. We detail this in Algorithm \ref{alg:profileprimitives}.

\vspace{-2em}
\subsection{Relationship Discovery (Column–Column Links)}
\begin{figure}[H]
  \centering
    \includegraphics[width=0.70\textwidth]{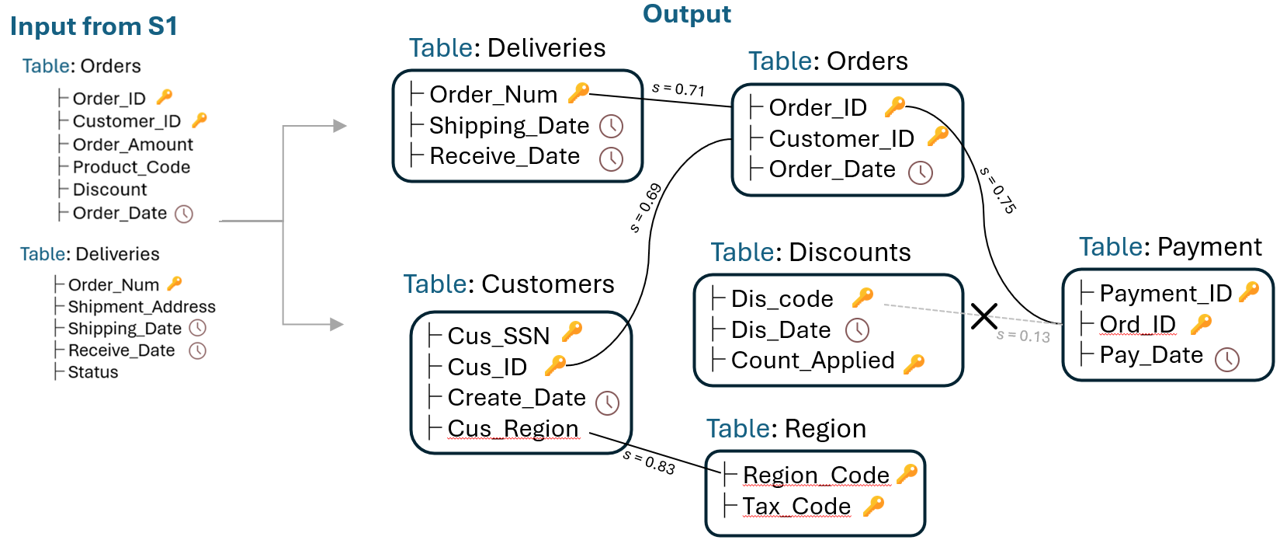}
    \caption{Illustration of column–column link discovery. Edges arise from high-overlap column pairs (approximate inclusion dependencies). Dashed edges indicate weaker type compatibility. Only pairs with final score $s \ge \tau_S$ are retained.}
    \label{fig:DiscoverLinks}
\end{figure}

We infer table connections by scoring every cross-table column pair  using a
\textit{type-aware similarity measure} designed to remain reliable under key sparsity (Figure~\ref{fig:DiscoverLinks}). For any pair consisting of column $c_i$ from 
table $T_i$ and column $c_k$ from table $T_k$, the score is
\[
s(c_i, c_k) = TC(c_i, c_k) \cdot Q(c_i, c_k) \cdot U(c_i, c_k) \cdot R(c_i, c_k),
\]

where each factor contributes a different aspect of link quality:
\begin{itemize}
    \item \textbf{Type compatibility $TC$}: downweights pairs whose data types are weakly compatible  
    (e.g., string vs.\ numeric, date vs.\ free text).
    \item \textbf{Column quality $Q$}: penalizes low-quality columns using completeness and distinctness  
    (\( Q = \min\{\gamma(c_i)u(c_i),\, \gamma(c_k)u(c_k)\} \)).
    \item \textbf{Coverage $U$}: ensures the overlap is not driven by rare or fringe values.
    \item \textbf{Value similarity $R$}: measures actual value alignment. We use Jaccard Similarity (JS) for symbolic columns
    and a JS/Kolmogorov-Smirnov (KS) blend for numeric distributions.
\end{itemize}

We retain all column pairs with $s(c_i, c_k) \ge \tau_S$ (optionally keeping only the top-$K$ per table) and form an undirected multigraph of candidate inter-table links. \ We denote the resulting set of retained column–column links as $\mathcal{L}$. Algorithm~\ref{alg:DiscoverLinks} provides full details. We also report our thresholds used in Table \ref{tab:hyperparameters}.

Because exact INDs often fail under key sparsity and partial overlaps, we rely on type-aware value similarity (Jaccard, JS/KS) to detect approximate table relationships that remain stable even when declared keys are missing or inconsistently populated.

% Explain each algorithm.
% Layman explaination not just referencing. 
% explain possibly by line numbers etc. Either comments or line number. 
% Algorithm can sit in appendix for referencing instead of in paper. 

\subsection{Temporalization and Sequencing}\label{sec:temporalization}
\begin{figure}[H]
  \centering
    \includegraphics[width=1\textwidth]{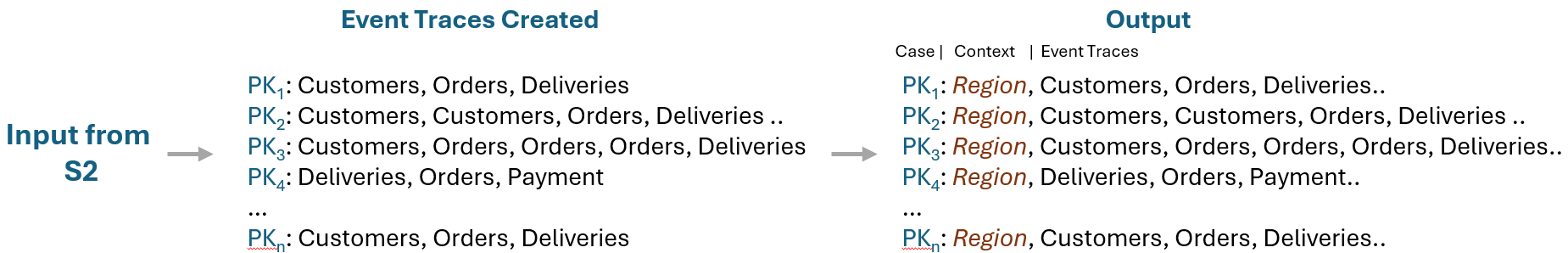}
    \caption{From Stage~2 links (left) we propagate sampled primary keys to collect timestamped rows, assemble per-key event traces (center), and emit case traces with optional context (right)}
    \label{fig:Sequencing}
\end{figure}

Using the retained link set~$\mathcal{L}$, we propagate sampled case identifiers across all reachable tables and collect their corresponding rows and non-null timestamp candidates (Figure~\ref{fig:Sequencing}). Each (row, timestamp-column) pair is treated as a provisional event.

Events for a case are sorted by timestamp to obtain a total order; ties are broken deterministically using table name and row identifier. Adjacent duplicate labels are collapsed\footnote{(e.g., A--A--B--C--C--C--D $\rightarrow$ A--B--C--D)}, while non-adjacent repetitions are preserved as they typically reflect genuine alternations. When timestamps coincide exactly, we randomize the within-timestamp order and average downstream statistics over five seeds. For efficiency, we subsample up to 100{,}000 identifiers per dataset (about 10\% in our experiments).

\paragraph{Optional context features.}
When business attributes (e.g., region, product, channel) are available, we embed them into a context vector~$c_s$ (dimension $d=16$) and concatenate any normalized numeric attributes. This vector is prepended to the event window passed to the TCN in Stage~4, conditioning precedence learning without changing the underlying event order.

Stage~3 outputs provenance-rich event traces in JSON format, exportable to XES or liftable to OCEL via additional identifier fields. Full pseudocode is provided in Algorithm~\ref{alg:assemblesequences}. This provides a lightweight conditioning signal for the TCN in \S3.4. 

\vspace{-3em}
\subsection{Sequence learning (ordering across tables)}
\begin{figure}[H]
  \centering
    \includegraphics[width=0.8\textwidth]{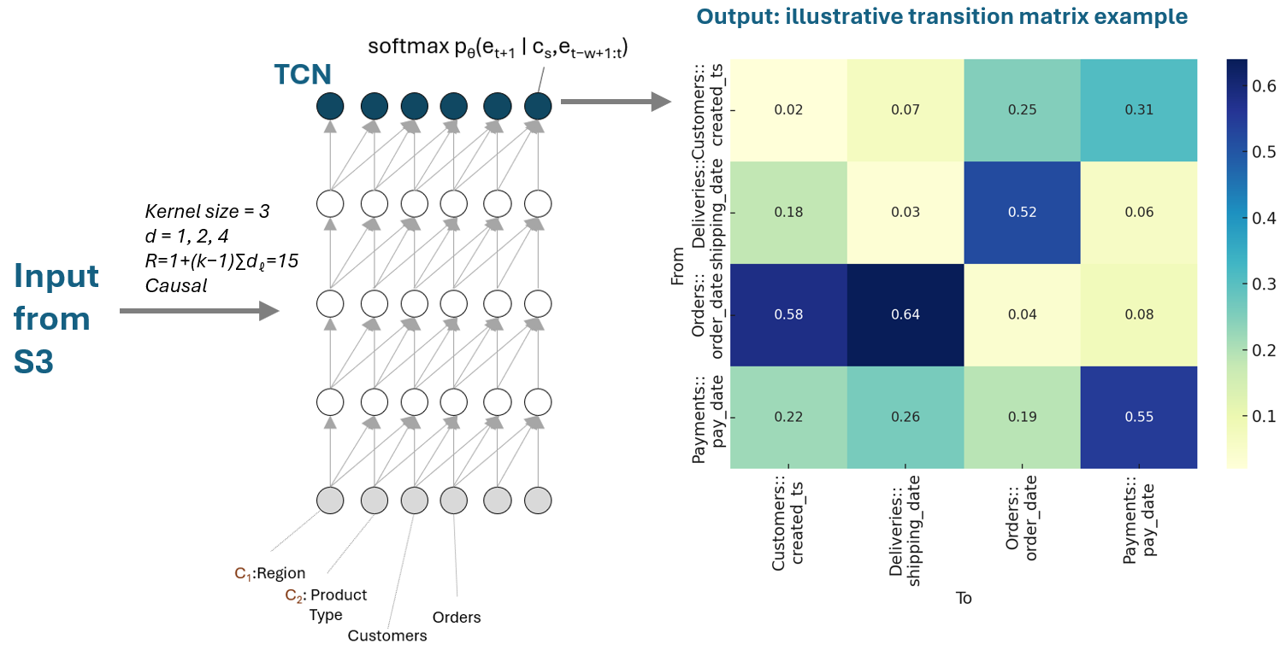}
    \caption{TCN-based next-event prediction and precedence extraction.}
    \label{fig:TCN Summary}
\end{figure}

In Stage~4, we learn cross-table precedence relations using a causal TCN that predicts the next event given a window of past events and an optional context vector (Figure~\ref{fig:TCN Summary}). Importantly, Stage~4 preserves the event sequences constructed in Stages~S1-S3 and learns precedence relations on top of them. The introduction of context-conditioned transition modelling sharpens precedence predictions. For a trace $e_{1:T}$, where $e_t \in V$ is the $t$-th event, the TCN receives the window
\[
e_{t-w+1:t} = (e_{t-w+1},\ldots,e_t),
\]

together with $c_s$, and estimates $p_\Theta(e_{t+1} \mid c_s, e_{t-w+1:t}).$
\vspace{-2.3em}
\paragraph{Aggregating next-event predictions.}
For any context set $C$, we form a transition matrix by averaging next-event probabilities over all prefixes whose last token is $i$:
\[
P_{ij}(C) = \frac{1}{|P_i(C)|} \sum_{(c_s,e_{t-w+1:t}) \in P_i(C)} p_\Theta(e_{t+1}=j \mid c_s,e_{t-w+1:t}),
\]
interpreting $P_{ij}(C)$ as the probability that $j$ follows $i$ under context $C$.

\paragraph{Extracting precedence and variability.}
We add a directed edge $i \rightarrow j$ when $P_{ij} - P_{ji} \ge \tau_\theta$. If $|P_{ij}-P_{ji}| < \tau_c$ and $\max(P_{ij},P_{ji}) \ge \tau_{\min}$, we mark the pair as exhibiting bidirectional order variability (BOV), i.e., both orders appear with similar frequency. Pairs that satisfy neither condition receive no precedence relation. Because our event logs contain only a single timestamp per event and no resource-level concurrency information, we report BOV-F1 rather than strict concurrency measures. A final transitive reduction yields a minimal precedence graph. Algorithm~\ref{alg:tcn} provides full details.

\section{Evaluation}
\subsection{Evaluation Framework}
For RQ1 we measure event recovery via position‑independent token accuracy (PITA) and per‑case event‑set precision/recall/F1; for RQ2 we assess ordering fidelity with edit distance, LCSr, and Kendall’s $\tau_b$, and precedence quality with edge precision/recall/F1 and BOV‑F1. These are described in Table \ref{tab:metrics_details}. For comparing sequence models we additionally report standard predictive metrics (perplexity, Top‑k, MRR, Brier). RQ3 uses the same metrics under controlled perturbations

\subsection{Data}
Publicly available datasets rarely preserve the relational and temporal structure needed for database-centric process mining \cite{Waibel2022CausalPM,10.1007/s10270-018-0664-7,d6b7e17c2bf7d0ee826d791398d6ac9fc7c78aff}. We therefore evaluate on five corpora spanning benchmarks, controlled synthetics, and industry data. Namely, TPC-H, TPC-E, Synth-Light, Synth-Heavy, and Industry-1 (Table~\ref{tab:datasets}).
\paragraph{Ethics and Threats to Validity.}
No human subjects or identifiable data are involved, and industry data are fully de-identified. Internal validity risks (e.g., leakage or seed sensitivity) are mitigated through temporal splits, fixed seeds, and early stopping. Construct validity concerns from imperfect ground truth are addressed via released mapping files, sensitivity checks, and multiple metrics. External validity is supported by evaluation across heterogeneous benchmarks, synthetics with induced drift, and one industry dataset. Statistical-conclusion validity is ensured via non-parametric tests with Holm correction and effect-size reporting.

\paragraph{Ground-truth logs and precedence.}
For all datasets, the data generator provides both databases and reference logs/precedence relations\footnote{Repository available at git clone \url{https://<ACCESS_TOKEN>@github.com/joellim2022/schema-agnostic-process-trace-construction.git}}. TPC-H/E data were generated using standard TPC tools. Table-level precedence is derived from the business narratives and schema documentation in the official specifications.\footnote{\url{https://www.tpc.org/tpc_documents_current_versions/current_specifications5.asp}} For Industry-1, we derive an initial precedence graph from system specifications and audit trails, validated independently by two domain experts.

\subsection{Baselines}

\paragraph{Deterministic Pipeline – Minimal Viable Proxy (DP-MVP).} We define a minimal deterministic surrogate to provide a lower bound for the add-in analysis. It 
(i) identifies keys and timestamps via simple name heuristics (e.g., \texttt{\_ID}, \texttt{\_NO}, \texttt{\_DATE}),  
(ii) links tables through name matches or high Jaccard overlap,  
(iii) orders events using a single global timestamp or table order, and  
(iv) derives precedence by the most frequent successor.

\paragraph{Deterministic Pipeline: S1--S3 (DP)}
DP applies Stages~S1--S3 exactly as defined but omits the learned precedence module (S4).  
It reconstructs traces deterministically and isolates the added contribution of learned ordering.

\paragraph{Sequence-learning baselines for S4.} To evaluate Stage~S4, we compare the TCN against a 3-gram language model and an LSTM trained on the same traces and tokenization. This isolates whether gains arise from the TCN architecture versus learned sequence modeling in general. Our ablation details are given in Table \ref{tab:surrogates}.

\paragraph{Schema-centric methods.}
Schema-centric techniques \cite{doi:10.1142/S021884301550001X,10.1007/978-3-642-36285-9_36} assume rediscoverable PK/FK constraints and a single timestamp attribute, or focus on key discovery rather than log or precedence construction.  
These assumptions do not hold in key-sparse, multi-timestamp datasets, and substantial manual configuration would be required. We therefore treat these as complementary methods rather than numerical baselines.

\subsection{Experimental Setup and Statistical Analysis}
All experiments use temporal hold-out splits, five random seeds, and shared tokenization. TCN hyperparameters are provided in Table~\ref{tab:tcn-config}. Optimizers, batch sizes, and hardware configurations (GPUs/CPUs, peak memory) are summarized in the Appendix.

For statistical testing, per-case metrics are averaged per dataset and compared against the deterministic baseline using the Wilcoxon signed-rank test with Holm correction. We report non-parametric effect sizes (Cliff’s~$d$) and 95\% BCa bootstrap confidence intervals from 10{,}000 resamples.

For reproducibility, we release scripts to regenerate datasets and results. Public and synthetic datasets are archived with DOIs, while industry results are reported only in aggregate form. For conformance experiments (§5.3), we discover models using Inductive Miner–infrequent ($\gamma=0.2$) and compute fitness/precision via ETConformance alignments.

\subsection{Analytical Design}

\paragraph{End-to-end comparison.}
We compare the deterministic pipeline (S1--S3) and the full pipeline (S1--S4) against baselines on event recovery (RQ1), ordering fidelity, and precedence accuracy (RQ2). For sequence learning, the TCN is contrasted with n-gram and LSTM baselines using the predictive metrics from RQ2. To isolate each stage’s contribution, we use (i) an add-in ladder that starts from DP-MVP and re-enables S1--S4 sequentially, and (ii) a knock-out analysis that replaces one stage at a time with its surrogate.

Importantly, S4 does not modify the event traces produced by S1--S3. Instead, S4 affects only the \emph{precedence-learning stage}: the deterministic variants (DP-MVP and DP) aggregate successor frequencies across all contexts, whereas the TCN learns \emph{context-conditioned} transition patterns (e.g., Equity vs Futures), resulting in sharper and more accurate precedence relations. Any observed differences in reconstruction-level metrics between DP and the full pipeline arise from aggregation over context partitions rather than changes to the underlying reconstructed traces. These experimental configurations are listed in Table~\ref{tab:surrogates} (Appendix).

\section{Results}
We organize the results around the research questions RQ1–RQ3 (§4.1), covering:
(i) end-to-end reconstruction quality of the extracted traces (RQ1),
(ii) the contribution of each pipeline stage to this quality and to precedence learning (RQ1, RQ2), and
(iii) robustness to schema/timestamp drift and noise or missing data (RQ3).

\begin{table}[t]
\centering
\scriptsize
\caption{Unified reconstruction and precedence-graph accuracy metrics across all datasets.}
\label{tab:unified-results}

\begin{tabular}{lcccccccccc}
\toprule
Dataset & PITA & ED & LCSr & $\tau_b$ & Prec & Rec & F1 & BOV-F1 & $\tau_\Theta$ & $\tau_c$ \\
\midrule
TPC-H       & 1.00 & 0    & 1.00 & 1.00 & 0.81 & 0.78 & 0.79 & 0.00 & 0.58 & 0.12 \\
TPC-E       & 0.88 & 27.4 & 0.82 & 0.75 & 0.84 & 0.80 & 0.82 & 0.77 & 0.60 & 0.14 \\
Synth-Light & 0.95 & 18.9 & 0.85 & 0.85 & 0.79 & 0.76 & 0.77 & 0.72 & 0.53 & 0.11 \\
Synth-Heavy & 0.82 & 31.1 & 0.72 & 0.63 & 0.75 & 0.73 & 0.74 & 0.69 & 0.47 & 0.08 \\
Industry-1  & 0.84 & 26.7 & 0.78 & 0.71 & 0.82 & 0.79 & 0.81 & 0.75 & 0.56 & 0.13 \\
% Industry-2  & 0.74 & 35.1 & 0.61 & 0.0.68 & 0.71 & 0.77 & 0.67 & 0.61 & 0.51 & 0.13 \\
\bottomrule
\end{tabular}
\end{table}

\subsection{RQ1 — Event Recovery (Presence and Accuracy)}
Across all corpora, Stages~S1-S3 achieve high event-set recovery, with PITA ranging from 0.82 (Synth-Heavy) to 1.00 (TPC-H) and strong ordering fidelity (ED, LCSr, $\tau_b$; Table~\ref{tab:unified-results}). Accuracy declines on TPC-E and Synth-Heavy, which exhibit the greatest key sparsity and timestamp ambiguity, but remains sufficient to provide reliable input for learned precedence in Stage~S4. The drop from PITA 0.95 (Synth‑Light) to 0.82 (Synth‑Heavy) mirrors the increase in key sparsity and overlapping timestamp semantics in the generator; in practice this means that under severe drift our pipeline still recovers most events but loses some intra‑table granularity.

\subsection{RQ1/RQ2 — Stage Attribution (What Each Stage Adds)}
Across all datasets, the add-in ladder (Table~\ref{tab:laddered}) shows monotonic gains in reconstruction and precedence metrics as S1-S4 are enabled. Knockout results (Table~\ref{tab:knockouts}) indicate that S2 contributes most to trace completeness and ordering, while replacing S4 with an n-gram model leaves reconstruction unchanged but reduces precedence F1 by 0.11.  The FK-only surrogate confirms that schema-level keys alone are insufficient in key-sparse, multi-timestamp environments. Overall, S1-S3 drive reconstruction quality, and S4 provides the learned cross-table ordering.
\vspace{-2.3em}
\begin{table}[h!]
\centering
\caption{Averaged Stepwise add-in across all Dataset}
\label{tab:laddered}
\renewcommand{\arraystretch}{1.2}
\setlength{\tabcolsep}{6pt}
\scriptsize
\begin{tabular}{lcccc}
\toprule
\textbf{Variant} & \textbf{PITA $\uparrow$} & \textbf{LCSr $\uparrow$} & \textbf{Kendall $\tau_b \uparrow$} & \textbf{Precedence F1 $\uparrow$} \\
\midrule
Deterministic Pipeline (MVP):   & ---   & ---   & ---   & ---   \\
+ S1 Profiling    & 0.52  & 0.46  & 0.36  & 0.66  \\
+ S2 Links        & 0.65  & 0.61  & 0.51  & 0.73  \\
+ S3 Temporalization & 0.72  & 0.69  & 0.59  & 0.78  \\
+ S4 TCN (\textbf{Full}) & 0.84  & 0.78  & 0.71  & 0.81  \\
\bottomrule
\end{tabular}
\end{table}

%The large gain from S2 suggests that investing effort in high‑quality similarity thresholds pays off more than tuning the TCN; in environments with limited compute, one could deploy S1–S3 alone and already obtain competitive traces.
\vspace{-2.5em}
\subsection{RQ2 — Learning to Order Across Systems}
The TCN outperforms n-gram and LSTM baselines on next-event prediction and uncertainty metrics (Table~\ref{tab:next_event}), and these predictive gains translate into higher conformance under identical mining and alignment settings. Precedence precision/recall/F1 (Table~\ref{tab:unified-results}) remain in the 0.74–0.82 range with concurrency-F1 between 0.69–0.77. Thus, learning precedence with the TCN improves both predictive quality and the fidelity of recovered precedence and concurrency relations.Although LSTMs perform competitively in our setting, we adopt TCNs because their convolutional structure is well known to handle long-range temporal patterns effectively, which is important for cross-table event sequencing.

\vspace{-2.2em}

\begin{table}[H]
\centering
\caption{Knockout from Full on Industry-1 (surrogate in parentheses).}
\label{tab:knockouts}
\renewcommand{\arraystretch}{1.1}
\setlength{\tabcolsep}{3pt}
\scriptsize
\begin{tabular}{@{}lccc p{0.40\textwidth}@{}}
\toprule
\textbf{Variant} & $\Delta$PITA & $\Delta$LCSr & $\Delta$Precedence F1 & \textbf{Interpretation} \\
\midrule
Full & --- & --- & --- & --- \\
-- S1 ($\rightarrow$ S1-Weak) 
  & -0.09 & -0.07 & -0.05 
  & More spurious links $\rightarrow$ extra/noisy events \\
-- S2 ($\rightarrow$ S2-FK-Only) 
  & -0.12 & -0.11 & -0.08 
  & Fewer true cross-table joins (key-sparsity hurts) \\
-- S3 ($\rightarrow$ S3-STGS) 
  & -0.06 & -0.09 & -0.07 
  & Single-time sorting loses intra-table event granularity \\
-- S4 ($\rightarrow$ n-Gram) 
  & -0.03 & -0.06 & -0.11 
  & Loses long-range/contextual precedence \\
\bottomrule
\end{tabular}
\end{table}

\vspace{-5em}

\begin{table}[H]
\centering
\caption{Next-event quality \& conformance (mean over 5 seeds, Synth-Heavy with 30\% Noise).}
\label{tab:next_event}
\scriptsize
\renewcommand{\arraystretch}{1.1}
\setlength{\tabcolsep}{3.5pt}  % <-- tighter columns
\begin{tabular}{lccccccc}
\toprule
Model & Perplx. $\downarrow$ & Top-1 $\uparrow$ & Top-3 $\uparrow$ & 
MRR $\uparrow$ & Brier $\downarrow$ & Fit. $\uparrow$ & Prec. $\uparrow$ \\
\midrule
\textit{n}-Gram ($k{=}3$) & 16000 & 0.08 & 0.149 & 0.258 & 0.118 & 0.63 & 0.07 \\
LSTM & 1.874 & 0.771 & 0.93 & 0.883 & 0.019 & 0.780 & 0.73 \\
\textbf{TCN (Full)} & \textbf{1.588} & \textbf{0.861} & \textbf{0.970} & 
\textbf{0.924} & \textbf{0.019} & \textbf{0.788} & \textbf{0.862} \\
\bottomrule
\end{tabular}
\end{table}

\subsection{RQ3 — Change Resilience and Robustness}
Under controlled drift on Synth-Heavy (Table~\ref{tab:drift_robustness}), the full pipeline retains over 80\% of its clean-data PITA and LCSr even under substantial attribute swaps and timestamp missingness, whereas the deterministic baseline drops below 50\%. All corpora are processed within our hardware budget. These results show that the pipeline remains viable for systems where schemas and designs evolve continuously.

\vspace{-2em}

\begin{table}[H]
\centering
\scriptsize
\begin{tabular}{lcccc}
\toprule
\multirow{2}{*}{Drift scenario (Synth-Heavy)} &
\multicolumn{2}{c}{Full pipeline} &
\multicolumn{2}{c}{DP baseline} \\
\cmidrule(lr){2-3}\cmidrule(lr){4-5}
 & PITA $\uparrow$ & LCSr $\uparrow$ & PITA $\uparrow$ & LCSr $\uparrow$ \\
\midrule
0\% attr swap, 0\% TS missing      & 0.82 & 0.72 & 0.70 & 0.55 \\
10\% attr swap, 5\% TS missing     & 0.77 & 0.68 & 0.52 & 0.41 \\
30\% attr swap, 15\% TS missing    & 0.68 & 0.65 & 0.34 & 0.27 \\
\midrule
Rel.\ to clean (worst drift)       & 0.83 & 0.83 & 0.49 & 0.49 \\
\bottomrule
\end{tabular}
\caption{Robustness to synthetic schema/timestamp drift on Synth-Heavy.
Each row applies controlled attribute renaming and timestamp missingness.
Under the highest drift level, the full pipeline retains about 83\% of clean-data
PITA and LCSr, while the deterministic DP baseline retains below 50\%.}
\label{tab:drift_robustness}
\end{table}

\vspace{-4.4em}
\section{Conclusion and Future Works}
\vspace{-0.1em}
We introduced a schema-agnostic framework for automatically constructing event logs directly from raw relational data. Our results challenge the long‑standing assumption in IS engineering that reliable event logs presuppose clean schemas and curated keys. By treating relational contents as the primary object, and learning linkages and precedence from data rather than from ERDs, it is possible to construct high‑quality logs even as schemas drift and systems evolve. This suggests a shift from schema‑first to data‑first log engineering in continuously changing IS environments. More broadly, it points to an IS engineering paradigm where operational schemas are allowed to evolve freely, while a data-driven log-engineering layer preserves continuity for analysis, auditing, and redesign. Across benchmark, synthetic, and industry datasets, our pipeline consistently reconstructs event sets, produces high-fidelity orderings, and learns accurate cross-table flow relations in key-sparse, multi-timestamp environments. The add-in and knock-out analyses further confirm that each stage contributes meaningfully to overall performance, validating our pipeline’s modular design.

\paragraph{Limitations and future work}  
Our approach assumes a basic level of data integrity; severe timestamp sparsity, homonymous attributes, or highly parallel workflows can make reconstruction ambiguous. The scarcity of public datasets with ground-truth relational processes also limits cross-study comparability. Identifier profiling currently considers columns in isolation and thus misses composite keys, although this is consistent with our schema-agnostic focus on quasi-identifiers rather than exact PK/FK recovery. Future work includes extending profiling to unique column combinations, stress-testing the pipeline under stronger noise and drift, automatically detecting contextual attributes (e.g., via clustering or LLM-based annotation), and refining sequence learning to better capture parallel and cyclic behavior.

\appendix
\setcounter{table}{0}
\counterwithin{table}{section}
\renewcommand{\thetable}{A.\arabic{table}} % A.1, A.2, etc.

\section*{Appendix}
\vspace{-1.3em}
\begingroup
\tiny
\setlength{\parskip}{1pt}
\setlength{\parindent}{6pt}
\setlength{\abovedisplayskip}{2pt}
\setlength{\belowdisplayskip}{2pt}
\captionsetup{font=tiny,aboveskip=2pt,belowskip=2pt}
\setlength{\tabcolsep}{1.5pt}
\renewcommand{\arraystretch}{0.92}

\begin{multicols}{2}
\begin{minipage}{\linewidth}
\captionof{algorithm}{\textsc{Profiling}}
\label{alg:profileprimitives}
\tiny
\begin{algorithmic}

\ForAll{$T_i \in T$}
  \ForAll{$c \in T_i$}

    % --- IDENTIFIER STEP ---
    \State Compute distinctness $u(c)$, completeness $\gamma(c)$, and length stability $\lambda(c)$
    \State Compute identifier score:
    \[
        s_{\mathrm{pk}}(c) \gets u(c)\,\gamma(c)\,\lambda(c)
    \]

    % --- TIMESTAMP STEP ---
    \State Compute basic timestamp feasibility:
    \State \hspace{1em} $\textsf{is\_parsable\_datetime}(c)$
    \State \hspace{1em} $\textsf{distinct\_time\_ratio}(c)$
    \State \hspace{1em} $\textsf{passes\_timestamp\_validity\_checks}(c)$

    \If{$\textsf{passes\_timestamp\_validity\_checks}(c)$}
        \State Compute temporal dispersion $(sd, IQR, H)$
    \EndIf

   \State Compute timestamp score: $s_{\mathrm{dt}}(c) $

  \EndFor

  \State $\mathrm{PK}(T_i) \gets \textsc{TopK}(s_{\mathrm{pk}})$
  \State $\mathrm{DT}(T_i) \gets \textsc{TopK}(s_{\mathrm{dt}})$

\EndFor

\State \Return $\{\mathrm{PK}(T_i), \mathrm{DT}(T_i)\}$
\end{algorithmic}
\end{minipage}

\vspace{2mm}

% ======================================================
%  Algorithm 2: DiscoverLinks
% ======================================================
\begin{minipage}{\linewidth}
\captionof{algorithm}{\textsc{DiscoverLinks}}
\label{alg:DiscoverLinks}
\tiny
\begin{algorithmic}
\State $\mathcal{L} \gets \varnothing$
\ForAll{$(c_i,c_k)$ from different tables}
    \If{$TC[\text{type}(c_i),\text{type}(c_k)] = 0$}
        \State \textbf{continue}
    \EndIf
    \State $Q \gets \min\{\gamma(c_i)u(c_i),\gamma(c_k)u(c_k)\}$
    \If{$V_i \cap V_k = \varnothing$}
        \State \textbf{continue}
    \EndIf
    \State $U\gets\min\big(|V_i\cap V_k|/|V_i|,\ |V_i\cap V_k|/|V_k|\big)$
    \If{$c_i,c_k$ numeric}
        \State $R \gets \text{JS/KS blend}$
    \Else
        \State $R \gets \text{Jaccard}(V_i,V_k)$
    \EndIf
    \State $s(c_{i},c_{k}) \gets TQU R$
    \If{$s(c_{i},c_{k})\ge\tau_S$}
        \State add $(c_i,c_k)$ to $\mathcal{L}$
    \EndIf
\EndFor
\State \Return $\mathcal{L}$
\end{algorithmic}
\end{minipage}

\vspace{2mm}

% ======================================================
%  Algorithm 3: AssembleSequences
% ======================================================
\begin{minipage}{\linewidth}
\captionsetup{font=tiny}
\captionof{algorithm}{\textsc{AssembleSequences}}
\label{alg:assemblesequences}
\tiny
\begin{algorithmic}

\ForAll{$s$ \textbf{in} sample}
  \State collect timestamped events across link set $R$
  \State $\varepsilon_s \gets \textsc{SortByTime}(\text{events})$
  \State serialize $\varepsilon_s$ to JSON
\EndFor

\State \Return $\{\varepsilon_s\}$

\end{algorithmic}
\end{minipage}

\vspace{2mm}

% ======================================================
%  Algorithm 4: TrainAndInferTCN
% ======================================================
\begin{minipage}{\linewidth}
\captionsetup{font=tiny}
\captionof{algorithm}{\textsc{TrainAndInferTCN}}
\label{alg:tcn}
\tiny
\begin{algorithmic}

\Require Sequences $\{\varepsilon^{c}_s\}$; window $w$; model $f_\Theta$; thresholds $\tau_\theta,\tau_c,\tau_{\min}$; vocabulary $\mathcal{V}$ incl.\ $\langle \mathrm{eos} \rangle$
\Ensure transition matrix $\mathbf{P}(\mathcal{C})$; precedence $\delta(\mathcal{C})$; concurrency flags

\State tokenize events (table::timefield) and contexts
\State append $\langle \mathrm{eos} \rangle$ to each trace

\State build training pairs $([\mathbf{c}_s; e_{t-w+1:t}], e_{t+1})$ with pre-padding to length $w$

\State train $f_\Theta$ using cross-entropy over causal dilated convolutions

\ForAll{$\mathcal{C}$}

    \State initialize $C_{ij}(\mathcal{C}) \gets 0$, \, $N_i(\mathcal{C}) \gets 0$ for all $i,j \in \mathcal{V}$

    \ForAll{prefixes in $\mathcal{C}$ with last token $i$}
        \State $\hat{\mathbf{p}} \gets f_\Theta([\mathbf{c}_s; e_{1:t}])$
        \State $C_{i\cdot}(\mathcal{C}) \gets C_{i\cdot}(\mathcal{C}) + \hat{\mathbf{p}}$
        \State $N_i(\mathcal{C}) \gets N_i(\mathcal{C}) + 1$
    \EndFor

    \ForAll{$i,j \in \mathcal{V}$}
        \State $P_{ij}(\mathcal{C}) \gets \dfrac{C_{ij}(\mathcal{C})}{\max(1, N_i(\mathcal{C}))}$
    \EndFor

    \ForAll{$i,j \in \mathcal{V}$}
        \State $\delta_{ij}(\mathcal{C}) \gets \mathrm{sign}_{\tau_\theta}(P_{ij}(\mathcal{C}) - P_{ji}(\mathcal{C}))$
    \EndFor

    \ForAll{$i,j \in \mathcal{V}$}
        \If{$|P_{ij} - P_{ji}| < \tau_c$ \textbf{and} $\max(P_{ij},P_{ji}) \ge \tau_{\min}$}
            \State flag $(i,j)$ as concurrent
        \EndIf
    \EndFor

\EndFor

\State \Return $\{\mathbf{P}(\mathcal{C}), \delta(\mathcal{C})\}$

\end{algorithmic}
\end{minipage}

\vspace{2mm}

% ======================================================
%  TABLES
% ======================================================

\begin{minipage}{\linewidth}
\captionsetup{font=tiny}
\captionof{table}{Hyperparameters.}
\label{tab:hyperparameters}
\tiny
\begin{tabular}{llc}
\toprule
Symbol & Description & Value \\
\midrule
$\tau_S$          & Similarity cutoff (S2)                    & .20 \\
$\tau_{\text{parse}}$  & Timestamp parse success rate (S1)        & 1\\
$\tau_\delta$      & Distinct-time ratio threshold (S1)       & .02 \\
$k_{\text{pk}}$     & Top-$k$ identifier-like columns          & 3 \\
$k_{\text{ts}}$     & Top-$k$ timestamp-like columns           & 2 \\
$B$                & Histogram bins for JS/KS                 & 100 \\
$\alpha$           & JS/KS mixture weight                     & .5 \\
$\tau_{\text{cov}}$ & Minimum row-coverage for column links    & .05 \\
$\tau_{\min}$       & Minimum support for precedence edges     & .05 \\
$\tau_\theta$       & Precedence asymmetry cutoff ($P_{ij}-P_{ji}$) & .10 \\
$\tau_c$           & Symmetry window for BOV classification   & .10 \\
\bottomrule
\end{tabular}
\end{minipage}

\begin{minipage}{\linewidth}
\captionsetup{font=tiny}
\captionof{table}{Similarity and precedence metrics.}
\label{tab:metrics_details}
\tiny

\begin{tabularx}{\linewidth}{lX}
\toprule
\textbf{Metric} & \textbf{Summary} \\
\midrule
PITA      & F1-like event multiset similarity where $PITA(A, B) = 2·|A \cap B| / (|A| + |B|)$\\
LCSr      & Normalized longest common subsequence $LCSr(A, B) = LCS(A, B) / max(|A|, |B|)$ \\
Prec.\ F1 & Strict $x \prec y$ order where Prec.F1 = F1 over pairs (x, y) where x < y in ground truth\\
BOV.\ F1 & F1 over bidirectional precedence: $x \prec y$ and $y \prec x$ \\
\bottomrule
\end{tabularx}

\end{minipage}

\begin{minipage}{\linewidth}
\captionsetup{font=tiny}
\captionof{table}{Dataset characteristics. ‘Tbl’ is the number of tables; ‘TS’ is the number of tables containing timestamp-like attributes; ‘Evts’ reports the median (50th percentile) and 90th-percentile number of events reconstructed per case.}
\label{tab:datasets}
\tiny
\begin{tabular}{lccccc}
\toprule
Dataset & Tbl & Rows & TS & Cases & Evts (50/90) \\
\midrule
TPC-H     & 8  & 866K      & 2  & 12k  & 2 / 2 \\
TPC-E     & 32 & $>$1.5M           & 8  & 18k  & 5 / 8 \\
Synth-L   & 20 & $>$100K          & 20 & 10k  & 8 / 12 \\
Synth-H   & 50 & $>$600K          & 46 & 25k  & 18 / 40 \\
Ind-1     & 27 & $>$500K          & 25 & 8k   & 15 / 28 \\
% Ind-2     & 62 & $>$3M          & 57 & 51k   & 18 / 33 \\
\bottomrule
\end{tabular}
\end{minipage}

% ---------------- TCN Config ----------------
\begin{minipage}{\linewidth}
\captionsetup{font=tiny}
\captionof{table}{TCN configuration used in all experiments.}
\label{tab:tcn-config}
\tiny
\begin{tabular}{ll}
\toprule
Component            & Setting            \\
\midrule
Input window length  & $w = 16$ events    \\
Label / ctx dims     & 16 / 16            \\
Blocks               & 3 residual         \\
Kernel size          & 3                  \\
Dilations            & (1, 2, 3)          \\
Hidden channels      & 64                 \\
Norm / activation    & LayerNorm / ReLU   \\
Dropout              & 0.1                \\
Loss / optimizer     & CE / Adam (1e-3)   \\
Batch size           & 64                 \\
Early stopping       & 10 epochs          \\
\bottomrule
\end{tabular}
\end{minipage}

% ---------------- Surrogates ----------------
\begin{minipage}{\linewidth}
\captionsetup{font=tiny}
\captionof{table}{Stage surrogates used in knockouts and DP-MVP.}
\label{tab:surrogates}
\tiny
\begin{tabular}{ll}
\toprule
Stage & Surrogate \\
\midrule
S1 Profiling       & S1-Weak: noisy profiling             \\
S2 Link discovery  & S2-FK-Only: declared/name FK         \\
S3 Temporalization & S3-STGS: single-TS global sort        \\
S4 Learning        & S4-n-Gram: $k=3$ model                \\
\bottomrule
\end{tabular}
\end{minipage}

\section*{Supplementary Materials}
The full source code and experimental scripts supporting this paper are provided through two access options:
Preferred:\begin{verbatim}
git clone https://github_pat_11B2UDGFQ0rMnEo12iSLrI_kbwy6WbYZm
8pdwHxAaFgx99fwUtFzU67ICFnI5QyL8PSABQBEOQNFFv6Zqp@github.com/j
oellim2022/schema-agnostic-process-trace-construction.git
\end{verbatim}
\begin{center}
Public Link: https://github.com/joellim2022/schema-agnostic-process-trace-construction
\end{center}
This public mirror will remain accessible for 30 days following the submission deadline.

\end{multicols}

\endgroup

\bibliographystyle{plain}
\bibliography{references}  % Ensure this points to your .bib file

\end{document}